 \documentstyle [12pt]  {article}
 \evensidemargin\oddsidemargin

\begin{document}

 \title{{\bf Commutative Fuzzy Geometry  and\\ Quantum
  Particle Dynamics \\ }}
 \author {{\bf S.N.Mayburov} \\
 Lebedev Inst. of Physics\\
   Leninsky Prospect 53,  Moscow, Russia, RU-117924\\
  e-mail: mayburov@sci.lebedev.ru \\}
 \date { }
 \maketitle
%


\begin{abstract}

 Commutative fuzzy geometry considered as  possible
  mathematical framework for reformulation
of quantum-mechanical  formalism in  geometric terms.
 In this approach,  states of massive
 particle $m$ correspond to  elements of fuzzy manifold
 called  fuzzy points. In 1-dimensional case, due to manifold
   specific (fuzzy) topology, $m$ space coordinate $x$ acquires
    principal uncertainty $\sigma_x$
and described by  positive, normalized density $w(x,t)$. Analogous
uncertainties appear for fuzzy point on $3$-dimensional manifold.
It's shown that $m$ states on such  manifold are equivalent to
vectors (rays) on complex Hilbert space, their evolution
correspond to Shroedinger dynamics of nonrelativistic quantum
particle.


\end{abstract}


\section{Introduction}

It's well-known that quantum mechanics (QM) can be consistently
described by several alternative  formalisms such as Shroedinger
(standard) one, algebraic QM, quantum logic, functional integral,
etc. \cite {Jauch}. In last years, possible reformulation of QM
formalism in geometric terms  is also extensively discussed
 \cite {Kib,Mar}. Really, in some fields of physics,
like optics and general relativity, geometric ideas have been
very useful, so one can hope that they
 also can help to study some important problems of quantum physics, first of all,
 quantum theory of gravity and   gauge fields.
Up to now, several alternative formalisms were proposed for
realization of QM geometrization: symplectic geometry, Hilbert
manifolds, Kahler bundles, etc.  (see \cite{Mar} and refs.
therein).
 In this paper, we'll study  approach to QM geometrization based on fuzzy calculus,
  in particular, it exploits fuzzy geometry formalism \cite {Dub,Gott}. During last
50 years, the fuzzy set theory and other branches of fuzzy
mathematics were applied in a wide range of scientific areas such
as biology, economics and computer science. From the early days
of its development, significant similarity of that theory and QM
formalism was noticed [6-8].  It was argued that the parameter and
proposition uncertainty
 (fuzziness), which is generic for fuzzy mathematics, can be
  equivalent to QM uncertainty of particle coordinate,
   momentum and other  observables \cite {Dod,Zee}.


It's notable that in modern QM the fuzzy (unsharp) observables
became important part of QM theory of measurements \cite
{Per},
hence QM formalism based on fuzzy methods can be interesting both
from fundamental and applied angles.
However, until now such studies were performed only in the context
  of fuzzy (multivalued) logics  (\cite {Pyk} and refs. therein).
   In distinction, our approach  deals with standard logics and
  based on the results of  fuzzy topology and geometry \cite {Zee,Pos,Sos}.
In particular, our formalism  exploits  system phase space
equipped with fuzzy topology; it's shown that the corresponding
structure of system state space is equivalent to QM Hilbert space
\cite {May2,May2+}. As the result, it permits to derive
quantum system evolution from geometric arguments. It's studied
here  for nonrelativistic particle, such system traditionally
exploited as testing ground for QM foundation studies. It's shown
that evolution of particle state corresponds to QM Shroedinger
dynamics \cite {May3, May4}. In this paper, we continue to study
relations between system states, its physical parameters and
underlying geometric structure. We consider here most general
situation, so the spatial state disjointness and nodal regions
will be accounted in our  formalism.

 In  mathematics,  formalism
exploited here  called fuzzy geometry \cite {Pos, Buck}, but in
modern physics such term ascribed to some noncommutative  field
theories \cite {Bal}, so to avoid  confusion,  it will be
called here commutative fuzzy geometry (CFG). Our paper organized
as follows. In section 2,  basic features of fuzzy topology and
geometry are reviewed and their relations with set theory and
topology are discussed.
 In section 3,  the model of 1-dimensional  particle evolution on CFG
 manifold considered and the resulting space of particle states
is derived. In section 4, particle evolution equation in this
model obtained. In section 5, general formalism for 3-dimensional
case constructed. Section 6 presents our concluding remarks.





 \section {Geometric Fuzzy Structures}

Here we'll consider  fuzzy structures important for formalism
construction, for the detailed review on CFG and related topics
see \cite {Pos,Sos,Buck}. The properties of fuzzy objects can be
introduced $ad\, hoc$, but it's worth to start from consideration
of their analogs in set theory which provides the useful link to
the realm of fuzzy structures and illustrates their physical
meaning.

Remind that in set theory  sets can be classified according their
ordering structure;
 the simplest case presents the  totally ordered  set, for all its element pairs
 $a_{k},a_{l}\,$    ordering relations $a_k\leq a_l$ (or vice versa)  fulfilled.
In distinction, for  partially ordered set (Poset),
 some its element pairs can
 obey to  incomparability  relations (IR) between
 them: \(a_j \sim a_k\). In this  case, both $a_j \le
a_k$ and $a_k \le a_j$ propositions
 are false \cite {Schr}; as will be shown, in some aspects IR
 is the discrete  analogue of fuzzy relations.
 To illustrate this analogy, consider
 poset $S^p=A^p \cup B$, which includes the
 subset of incomparable elements $A^p=\{ a_j \} $,
and ordered subset $B=\{b_i \}$.
 For the simplicity suppose that
in $B$ the element indexes grow correspondingly to their
ordering, so that $\forall \, i$,  $b_i \le b_{i+1}$;  an arbitrary pairs $a_j, a_i$ or $a_i, b_l$ can be
ordered, as well as incomparable.
Let's  consider open $B$ interval  $\{ b_{l},b_n\}$ with $l+2
\leq n$,
 and suppose that  $A^p$ element $a_j$ is confined in
  $\{b_{l},b_n\}$, i.e.
    $\forall k$; $\, k \le l \,$:
 $b_{k} \le a_j \,$ ; $\forall m$; $\, n \le m$ : $\,a_j \le b_{m}$,
 and  simultaneously  $a_j$ is incomparable with all other (internal) $\{b_{l}, b_n\} $
 elements: $b_i \sim a_j;\,$
   $\forall \, i \,; \, l+1\le i \le n-1 $. In this
case, $a_j$  is, in a sense, smeared over $\{b_{l}, b_n\} $
interval,  this is  analogue of $a_j$ coordinate uncertainty,
if to regard the
 sequence of $B$ elements $\{b_i \}$  as the discrete coordinate axe.
Plainly, $A^p, B$ element relations can be also described by the
binary matrix $M_{ij}$, such that $M_{ij}=0$ if $a_i, b_j$ are
ordered, $M_{ij}=1$ otherwise.


Next step in transition to fuzzy structures is to change the
set-theoretical relations between $S^p$ elements  to  fuzzy
relations. To perform it, in place of $M_{ij}$ one should put in
correspondence to each $a_j, b_i$ pair of $S^p$ set the
nonnegative, normalized weight  function $w^j_i \ge 0$
 with  norm $\sum_i w^j_i=1$.
In fuzzy set theory, $w^j_i$ characterizes the rate of closeness
(membership) between $a_j, b_i$ \cite {Dub,Gott}. In particular,
analogously to $M_{ij}$, $w^j_i=0$ means that $a_j, b_i$ are
ordered relative to each other, i. e. their closeness is null,
and if $w^j_i=1$ they are equal.
 For the example considered above, one can ascribe arbitrarily:
$w^j_i=(n-l-1)^{-1}$
to all $b_i$ inside $\{b_{l}, b_n\} $ interval,  $w^j_i=0$ for
other $b_i$, the interval width   called  tolerance scale \cite
{Zee}. In principle,  fuzzy relations can be introduced $ad\,
hoc$ without any referring to  set partial ordering, but it's
worth to start from considering  their analogy .

Similar structure can be introduced for the  set of continuum
power. As the example, consider the set
 $S^f=A^p \cup X$ where $A^p$ is the
same discrete subset,
$X$ is  continuous ordered subset. In this case, $A^p$ element
$a_i$ can be incomparable to some $X$ elements $\{x_u\}$, in
particular, such $x_u$ can constitute the interval on $X$. If the
flat metrics
 $\cal M$$(x,x')$ is defined on $X$,
 then it's
 equivalent to R$^1$  real number axe. Then,  fuzzy relations between elements $a_j,x$ are described
by real, nonnegative functions
 $w^j(x)\ge 0$ with  norm $\int w^j dx=1$ \cite {Gott}.
  $\{a_j\}$  called  fuzzy numbers $\tilde{x}_j$
  or in geometric framework,  $1$-dimensional fuzzy points (FPs);
$S^f$ is  fuzzy manifold denoted $\tilde{\rm{R}}^1$.
 \cite {Dub,
 Gott,Buck}.  Plainly, ordered point $x_c \in X$ is
characterized by $w^c(x)=\delta(x-x_c)$, hence the ordered points
and FPs can be regarded formally on the same ground.   Note that
in fuzzy mathematics
  alternative FP definitions also are exploited, we use here  one
 given in \cite {Gott,Buck}.
In 3-dimensional case,  one can consider  fundamental set
 $S^f_3=A^p \cup X'$ where $A^p$ defined above,
 $X'$ is  continuous set. Suppose that on $X'$ $3-$dimensional
 linear space R$^3$ with flat metrics $\cal{M}$$_{ij}$ is defined.
Then FP $a_j$  is
  described by  nonnegative function $w^j(\vec{r})$ with norm $\int w^j
  d^3r=1$, such  structure on $S^f_3$ elements constitutes  fuzzy manifold denoted as
$\tilde{\rm{R}}^3$ \cite {Gott,Buck}.

\section{Particle States on Fuzzy Manifold  }

In this section, we'll consider the model of particle evolution on fuzzy
manifold called fuzzy mechanics (FM); it will be constructed here as
the minimal theory,
 i.e. at every step we'll choose  ansatz with minimal number of theory parameters
 and its degrees of freedom (DFs).
 Such approach seems appropriate for QM reconstruction,
 since QM formalism contains only one theory parameter -  Plank constant
$\hbar$. We'll suppose also that FM possesses
  space and time transalation invariance and rotational invariance.
 In  classical mechanics, the particle described as
 material point $ {\vec{r}_a}(t) \in \rm{R}^3$, whereas in FM
formalism the particle $m$
 corresponds to  FP $a(t)$ on fuzzy manifold
and characterized by normalized positive density $w(\vec{r},t)$
on R$^3$. Beside $w$, $m$  state $|\zeta (t)\}$ called  the
fuzzy state, can  depend, in principle, on
  other $m$ DFs.

 We consider first FM construction for 1-dimensional manifold,
  because in this case, the theory premises are most simple and
transparent.
 Let's suppose that $m$ state is prepared at some $t_0$
  and consider $m$ average velocity on R$^1$
\begin {equation}
\bar{v}(t)= \frac{\partial}{\partial
t}\int\limits^{\infty}_{-\infty}xw(x,t)dx=
\int\limits^{\infty}_{-\infty}x \frac{\partial w}{\partial
t}dx                                    \label {AB}
\end {equation}
It's reasonable to assume that in general $\bar v(t)$ can be
independent of $w(x,t)$; we shall  look for additional $m$
 DFs in form of real functions  $q_j(x,t); j=1,...,n$ related by some algebbra.
Let's suppose that in FM $m$ state evolution
 is local, in particular,
 \begin {equation}
           \frac{\partial w}{\partial t}(x,t)=\Phi[w(x,t), q_1(x,t),...,q_n(x,t)]     \label {AXY}
 \end {equation}
where $\Phi$ is an arbitrary  function.
From $w$ norm conservation it follows that
 \begin {equation}
   \int\limits^{\infty}_{-\infty} \Phi(x,t)dx = \int\limits^{\infty}_{-\infty} \frac{\partial w}{\partial t}(x,t)dx=
       \frac{\partial }{\partial t}\int\limits^{\infty}_{-\infty}
       w(x,t)dx = 0
                         \label {AZZZ}
 \end {equation}
 One can substitute: $\Phi=-\partial_x J$,
here $J(x)$ is continuous, differentiable function,
 which obeys to the condition
 \begin {equation}
 J(\infty,t)-J(-\infty,t)=0    \label {X}
 \end {equation}
Really, any normalized, differentiable $w(x) \to 0$ for $|x| \to
\infty$; in this case, from R$^1$ reflection invariance it follows
that $J(\pm\infty)=0$.
 Analogously to fluid dynamics, $J$ can be decomposed formally as
$$
    J(x)=w(x)v(x)
$$
so that $v(x)$  corresponds to $1$-dimensional $w$ flow
 velocity.
 In these terms eq. (\ref{AXY})
can be rewritten in form of flow continuity  equation \cite {Lan}
 \begin {equation}
                  {\partial_t w}= -v  \partial_x w
                   - w {\partial_x v}    \label {AZ2}
 \end {equation}
We'll assume that $v(x,t)$ can be considered as independent $m$
DF, and $w,v$ functions possess the following continuity
properties (CP):
 $w, v\in \rm{C}^3(\rm{R}^1\times[\it{t_0,T}])$;
here $[t_0, T]$ is closed interval.
 The twople $\varrho^o=\{w, v\}$ called the
 observational $|\zeta\}$ representation, at this stage, it's
  just the list of independent $m$ DFs, their algebra is undefined at the moment.
To make the formalism consistent, beside  eq.
(\ref{AZ2}), which describes  $w$ evolution,  it's necessary to
find also
  equation for $v$ evolution.
 However, $\varrho^0$ ansatz isn't optimal for that purpose, so
it's instructive to seek alternative dynamical  $|\zeta\}$
representation $\eta$, for which it will become more simple and
straightforward.
 We'll suppose that such $\eta$ corresponds
  to the set of real functions $\{\eta_i(x)\}, \,
i=1,n_a$, plus some $\eta_i$ algebra.
 If this is the case, the most general
$\eta$ ansatz is $\eta_j(x)=\Upsilon^j_x(w,v)$ where
$\Upsilon^j_x$ are some $w, v$ functionals and $x$ is their
parameter. For $|\zeta\}$ characterized by two DFs
 $w, v$,  it's natural to start  $\eta$
ansatz search from complex $\eta(x)$, not assuming yet that
$\eta$ is $\rm{L}^2$-normalized.
Hence $n_a=2$ and $\eta(x)$ can be expressed as
 \begin {equation}
   \eta(x)= \Upsilon^1_x +i\Upsilon^2_x  =\Omega_x(w,v) e^{i\lambda_x(w,v)}
                                              \label {AY15}
 \end {equation}
 where
 $\Omega_x,\,\lambda_x$ are real functionals. It's natural to
assume also that if  $w(x_a)= 0$ for some $x_a$, then
$\eta(x_a)=0$ and vice versa. It can be easily shown that
 the corresponding minimal  ansatz  is $\Omega_x=f[w(x)]$ where  $f \to 0$
 for $w \to 0$.

 In addition, we'll suppose that
 in FM the particle $m$ possesses some  holistic properties,
  namely, its evolution can be characterized
  also by  particle velocity 'as the whole' $u(t)$ described by
the corresponding normalized distribution $w_{u}(u,t)$. Plainly,
such $u$ can be also considered as fuzzy value.
 In general, $w_u$ characterizes instant $w(x,t)$ variations, in
particular, the shift of its centre of gravity
  $ \bar{x}(t)$
 and variation of $w$ half-width (r.m.s.) $\sigma_x(t)$. It follows then
\begin {equation}
   \bar{u}= \int\limits^{\infty}_{-\infty} u w_u(u)du = \bar{v}=
 \int\limits^{\infty}_{-\infty} v(x) w(x) dx
                                     \label {AY5}
 \end {equation}
  Since $\bar{u}=\bar{v}$, we shall not
  assume
beforehand that $u$ is independent $m$ DF.
In place of  $u$, below it will be convenient to use   variable
  $p=\mu u$ where
  $\mu$ is  theory parameter; its distribution denoted
  $w_p(p)$.
 $m$ state $|\zeta\}$ presumably  contains the information on the expectation value
of any $m$ observable $Q$ in form of some $\eta$ functional.
In particular, $w_p(p)=F_p(\eta)$ and it can be shown
 that $F_p$ functional is related to $\eta(x)$ Fourier transform.
To prove it and calculate $\eta$, $w_p$,
  let's introduce  $\rm L^2$-normalized function
  $ \varphi(p)=w_p^{\frac{1}{2}}\exp(i\beta)$, here $\beta(p)$
is the auxilary  real function, on which  final $w_p$ ansatz
 wouldn't depend.
We shall look for $w_p$,  $\beta$ such that $\eta$ Fourier
decomposition on $X$ is equal to $\varphi$, i.e.
\begin {equation}
\varphi(p)  = \int\limits^{\infty}_{-\infty}\eta(x)e^{-ipx} dx=
\int\limits^{\infty}_{-\infty}
 f(w) e^{i\lambda_x-ipx} dx
                                                          \label{AQ}
\end {equation}
$w_p$ is normalized, so the application of Plancherele identity to
that norm gives
 \begin {equation}
\int\limits^{\infty}_{-\infty}w_p(p)dp  =
\int\limits^{\infty}_{-\infty} \varphi(p) \varphi^*(p)dp=
\int\limits^{\infty}_{-\infty} f^2(w)dx =1 \label {AY25}
 \end {equation}
  To calculate $f$, let's define the function $\Theta=f^2$
   and consider its variation $\delta\Theta$. As follows from  eq. (\ref{AY25}):
\begin {equation}
  \int\limits^{\infty}_{-\infty}\delta\Theta dx =
\int\limits^{\infty}_{-\infty} \frac{\partial \Theta} {\partial
w} \delta w dx =0
                                                         \label {AY333}
 \end {equation}
with additional $\delta w$ constraint: $\int \delta w dx=0$.
Let's substitute $\delta w={\partial_x \varpi}$; under these
conditions it's possible to choose $\varpi(\pm\infty)=0$, then
such $\varpi$ satisfies to the conditions of Du Bois-Reymond
lemma treated in appendix,
 see also  \cite {Var}. Its application gives: $\Theta=w$,
so that $f=\pm w^{\frac{1}{2}}$.
 Now $\bar p$ can be  calculated anew  from
 derivative Fourier transform  \cite {Rees}
\begin {equation}
\bar{p}=\int\limits^{\infty}_{-\infty} p \varphi(p)
\varphi^*(p)dp= -i\int\limits^{\infty}_{-\infty}\eta^* \frac{\partial
\eta}{\partial x}dx=\int\limits^{\infty}_{-\infty} \frac{\partial
\lambda_x}{\partial x} f^2 dx = \int\limits^{\infty}_{-\infty}
\frac{\partial \lambda_x}{\partial x} w dx
                   \label {AY95}
\end {equation}
From its comparison with eq. (\ref {AY5}) and equality
$\bar{p}=\mu \bar u$, it follows that if $v, w$ are independent DFs
then
\begin {equation}
   v(x)=\frac{1}{\mu}\frac{\partial \lambda_x}{\partial
x}(w,v)      \label {AY88}
\end {equation}
so that $\lambda_x$ is independent of $w$.
For our model CP,  it follows that
 $\lambda_x(w, v)= \gamma(x)$ where $\gamma$ is the functional:
 \begin {equation}
       \gamma(x)= \mu\int\limits^{x}_{-\infty} v(\xi) d\xi +c_{\gamma}  \label {AY}
 \end {equation}
 here  $c_{\gamma}$ is an arbitrary real number.
The resulting  $m$ state $|\zeta\}$ in $x$-representation is
equal to:
\begin {equation}
   \eta (x)=w^{\frac{1}{2}}(x)e^{i\gamma} \label {AYA2}
\end {equation}
 so $\eta$ is $\rm{L}^2$-normalized vector (ray) of complex Hilbert space $\cal
 H$, and it describes the  complete set of $m$ pure states.
 $w_p(p)$ and $\beta(p)$
 can be calculated  from eq. (\ref {AQ}) as functions of
 $w, \gamma$.
In particular,
\begin {equation}
       w_p(p)= |\int\limits^{\infty}_{-\infty} w^{\frac{1}{2}} e^{i\gamma-ipx} dx|^2  \label {AYA}
 \end {equation}
is independent of $\beta({p})$,  so $w_p$ is just $\eta$, i.e.
$w, v$ functional, the same is true for $\beta$.



\section{Linear Model of Fuzzy Dynamics  }


Now let's consider $m$ state evolution for obtained $\eta$ ansatz.
If $m$ DFs obey to model CP given in sect. 3, $ v(x,t)$ can be
treated  as
 $\gamma(x,t)$ derivative, assuming that $\gamma\in
 \rm{C}^4(\rm{R}^1\times[\it{t_0,T}])$; then,
  $\gamma$
can be used for $\eta (x,t)$ description on equal terms.
The evolution equation for $\eta$ presumably is of the first order
in time,
 i.e.
  \begin {equation}
                  i\frac{\partial \eta}{\partial t}= \hat{H} \eta.     \label {AZ4}
 \end {equation}
 In general, $\hat H$ can be  nonlinear operator, for the
 simplicity
 we shall start with the linear $\hat H$ and consider nonlinear case
later. Free $m$ evolution is invariant relative to $x$ space
shift on arbitrary $x_0$  performed by the operator
  $\hat{W}(x_0)=\exp({x_0\frac{\partial}{ \partial x}})$.
 Because of it,  corresponding operator $\hat{H}_0$ should commute with
$\hat{W}(x_0)$ for the arbitrary $x_0$, i.e.
$[\hat{H}_0,\frac{\partial}{ \partial x} ]=0$. It holds only if
$\hat {H}_0$ is differential polinom
 of the form
 \begin {equation}
   \hat {H}_0=
    - \sum\limits_{l=1}^n b_{2l} \frac{\partial^{2l} }{\partial x^{2l}}       \label {AU}
 \end {equation}
 where  $ b_{2l}$ are  arbitrary real values, $n$ is arbitrary number.
 From correspondence to classical mechanics, it supposed
that the influence of   potential field $U$ on $m$ evolution can
be accounted in $\hat H$ additively:
\begin {equation}
\hat {H}= \hat {H}_0+U(x,t)  \label {AVA}
\end {equation}
 where $U$ is real, differentiable
function. Now eq. (\ref {AZ4}) can be rewritten as
\begin {equation}
                  i\frac{\partial g}{\partial t}=
(i\frac{\partial w^{\frac{1}{2}}}{\partial t}
  - w^{\frac{1}{2}} \frac{\partial \gamma}{\partial t})e^{i\gamma}=
   e^{i\gamma}\hat{Z}g      \label {AV}
 \end {equation}
where $\hat{Z} = \exp(-i\gamma)\hat{H}$. Hence
\begin {equation}
\frac{\partial w^{\frac{1}{2}}}{\partial t}= \rm{im} \it
(\hat{Z}g) \label
                                                                 {AY85}
 \end {equation}
If to substitute $v(x)$ by $\gamma(x)$ in eq. (\ref {AZ2}) and
transform it to ${w}^{\frac{1}{2}}$ time derivative, it gives
\begin {equation}
\frac{\partial w^{\frac{1}{2}}}{\partial t}=
 -\frac{1}{\mu}  \frac{\partial
w^{\frac{1}{2}}}{\partial x}\frac{\partial \gamma}{\partial x}
 - \frac{1}{2\mu} w^{\frac{1}{2}} \frac{\partial^2 \gamma}{\partial
 x^2}     \label {BV}
\end {equation}
 Plainly,
the right parts of equations (\ref {AY85}) and (\ref {BV})
  should coincide for arbitrary $w, \gamma$, otherwise $\hat H$ ansatz would
  be incompatible with $w$ flow continuity described by eqs. (\ref{AZ2}, \ref{BV}).
Really, if they  differ, it will impose the additional constraint
on $w,\gamma$ of the form $\hat{L}_x(w,\gamma)=0$, where
$\hat{L}_x$ is the operator in $x$-derivatives. However, $w,
\gamma$ presumably are independent DFs, hence such constraint
would result in the open contradiction. Therefore,
 $\hat H$ ansatz can be obtained from the term by term comparison
 of eqs. (\ref{AY85}, \ref{BV}).
  In particular,  the equality of highest
  $\gamma$ derivative for $\rm{im}(\it{\hat{Z}g})$ and of left part of  eq. (\ref{BV}) gives:
 $$
 - b_{2l} w^{\frac{1}{2}} \frac{\partial^{2l} \gamma }{\partial
 x^{2l}}= -\frac{1}{2\mu}w^{\frac{1}{2}}\frac{\partial^2 \gamma }{\partial x^2}
 $$
  It follows that $b_2={1}/{2\mu}$ and $b_{2l}=0$ for  $l\geq 2$,  only in this case both
   expressions  for ${\partial_t w^{\frac{1}{2}}}$ would coincide.
  Therefore, $\eta$ free evolution is described by single
  term
  $$
        \hat{H}_0 = -\frac{1}{2\mu} \frac{\partial^2}{\partial x^2}
  $$
   and so $\hat H$
  of eq. (\ref {AVA})  is  Schroedinger Hamiltonian
  for  particle with mass $\mu$. $\gamma$ evolution equation
  can be extracted from eq. (\ref{AV}),  $v$ evolution equation  follows from it. Thus,
   the system of equations is obtained for $w(x,t), v(x,t)$  functions, which obey to our model
   CP, under these conditions it's equivalent to Schroedinger equation for
   $\eta(x,t)$,
   at least for $\eta\in \rm{C}^4(\rm{R}^1\times[\it{t_0,T}])$.
   Plainly, such description can be extended on $m$ states
   localized in open interval $\{x_1, x_2\}$ on $\rm R^1$ - 'particle in 1-dimensional box'.

 It was assumed
previously that $v(x)$ is independent  $m$ DF, but proposed
formalism is compatible also with hypothesis that $v$ is function
of some other $m$ state parameter
 independent of $w$. It was shown that for our model CP,
$v(x)$ is proportional to $\gamma(x)$ derivative,
hence it's possible to choose $\gamma$ as such fundamental $m$ DF,
such choice will be shown to possess important advantages.
For such ansatz, eq. (\ref{AY88}) doesn't generally define
$\gamma$,
 only inverse relation (\ref{AY}) holds and defines $v$. Thereon,
modified $m$ state observational representation becomes:
$\varrho^o_{\gamma}=\{w(x,t), \gamma(x,t)\}$,  in the same time,
$\eta(x,t)$ can be retained as $m$ state dynamical
representation. In our initial approach, $m$ state evolution
should be described by evolution equations for $w$ and $v$, but
now they become equivalent to  evolution equation for $\eta$,
which unambiguously defines $w$ and defines $\gamma$ up to
constant $c_{\gamma}$, from that $v$ is also unambiguously
defined. Hence it's possible to drop initial CP for $w, v$ for
$m$ states and exploit  CP for $\eta$ functions: $\eta\in
\rm{C}^2(\rm{R}^1\times[\it{t_0,T}])$. It's notable that such
evolution ansatz can be extended on $m$ states, localized
temporarily or permanently
 in several disjoint $\rm R^1$ regions \cite {Vlad}.
 On the opposite, for fundamental
DFs $w, v$
 unambiguous description of such disjoint state evolution  is impossible \cite {Wall}.


In this framework, the observable $p$ corresponds to the operator
$\hat{p}=-i {\partial_x}$ acting on $\eta(x)$.
 Thus, $x$ and $p$
 observables are described by the linear self-adjoint operators,
which obey to the commutation relation $[\hat{x},\hat{p}]=i$.
  We'll admit that in general   $m$ projective
observables $\{ Q_i\}$ correspond to linear, self-adjoint
operators on $\cal H$ with standard extension to  POVM \cite
{Per}. Probabilstic mixture of $m$ states described by density
matrix on $\cal H$.

\section { General Fuzzy Dynamics }

  $3$-dimensional FM, in fact, doesn't demand
 the serious modification of described formalism,
however, some new features appear. Beside topological distinctions
between $\rm R^1$ and R$^3$ spaces, they are related also to novel
state disjointness induced by nodes and noding regions in $R^3$.
In this case, monodromy of evolution equations should be accounted
properly.
 Below the formalism derivation is described briefly
making the impact on that new moments.
 On $\tilde{\rm{R}}^3$ manifold the particle $m$ corresponds to FP $a(t)$ characterized by
 its state $|\zeta(t)\}$. As was shown in sect. 3, it depends on density $w(\vec{r},t)$
 and some other $m$ DFs $\{q_i (\vec{r},t)\}$ which supposed to be the
  real functions.
 Assuming that $w$ evolution depends
  on local parameters only,
 it can be expressed as:
 \begin {equation}
 \frac{\partial w}{\partial t}(\vec{r},t)=
 -\Phi[w(\vec{r},t),q_1(\vec{r},t),...q_k(\vec{r},t)]     \label {AXYYY3}
 \end {equation}
 where $\Phi$ is an arbitrary  function. Then from $w$ norm
 conservation
 \begin {equation}
   \int \Phi (\vec{r},t)d^3r =
   \int \frac{\partial w}{\partial t}(\vec{r},t)d^3r=
       \frac{\partial }{\partial t}\int
       w(\vec{r},t)d^3r = 0
                         \label {AZZZZ3}
 \end {equation}
 where integration performed over R$^3$.
 Substituting $\Phi=-\rm{div} \it\vec{J}$ where $\vec{J}(\vec{r},t)$
  is differentiable function, it follows that
  $w$ evolution described by the flow continuity equation
\begin {equation}
                  \frac{\partial w}{\partial t}=-\rm{div}\it\vec{J}     \label {AXY5}
 \end {equation}
 assuming that $\vec{J}$ flow through surrounding infinite spherical surface $S$ is
 zero:
$$
   \int\limits_S\vec{J}d\vec{s}=0
$$
Really, any normalized, differentiable $w(\vec{r}) \to 0$ for
$|\vec{r}| \to \infty$, hence for any $\vec{r}_a$ such that
$|\vec{r}_a| \to \infty$, it should be $\vec{J}(\vec{r}_a) \to
0$, otherwise it would violate rotational invariance.
 One
can decompose formally $\vec{J}=w\vec{v}$, so that $\vec{v}$
corresponds to $w$ flow velocity. We  suppose that
$\vec{v}(\vec{r})$ is independent $m$ DF and $w,\vec{v}$  CP  are:
 $w(\vec{r},t), \vec{v}(\vec{r},t)\in \rm{C}^3(\rm{R}^3\times[\it{t_0,T}])$.
 $m$ state $|\zeta\}$ supposedly depends on  $w, \vec {v}$ DFs only, the
twople $\varrho^o=\{w, \vec{v}\}$ describes observational
$|\zeta\}$ representation on R$^3$.


 Analogously to 1-dimensional case, one should find for R$^3$ the dynamical $|\zeta\}$
 representation $\eta$ which supposedly described by  set
  of real functions $\{\eta_i(\vec{r})\}$, their algebra is undefined at this  stage;
 it can correspond, for example, to  quaternion. Yet, we'll
 start its search from  minimal $\eta$ ansatz
in form of  complex $w, \vec{v}$ functional
$\eta(\vec{r})=\Upsilon_{\vec{r}}(w,\vec{v})$.  It follows that
its minimal ansatz is
 $\eta=f(w)\exp[{i\Lambda_{\vec{r}}(w,\vec{v}) }]$ where $f$ is real function,
   $\Lambda_{\vec{r}}(w,\vec{v})$ is real functional and
$f(w) \to 0$ for $w\to 0$.
  It assumed also that $m$ state characterized  by  particle
  velocity 'as the whole'
 $ {\vec u}(t)$ with corresponding distribution  $w_u(\vec {u})$, so that
\begin {equation}
\langle \vec{u} \rangle= \int \vec{u} w_u(\vec{u}) d^3u= \langle
\vec{v} \rangle= \int \vec{v}(\vec{r})w(\vec{r})
d^3r                    \label {AX25}
 \end {equation}
$m$ fuzzy momentum defined as: $\vec{p}=\mu \vec{u}$
with normalized distribution $w_s(\vec{p})$.
Analogously to sect. 3, to  calculate $\eta$, $w_s$,
  one can introduce  $\rm L^2$-normalized function
  $ \varphi(\vec{p})=w_{s}^{\frac{1}{2}}\exp(i\beta)$, here $\beta(\vec{p})$
is the auxilary  real function; it supposed that $\varphi$ is
$\eta$ Fourier transform, i.e.
\begin {equation}
\varphi(\vec{p})  = \int\eta(\vec{r})\exp({-i\vec{p}\vec{r}})d^3r=
\int f(w) \exp({i\Lambda_{\vec r}-i\vec{p}\vec{r}}) d^3r
                                                      \label{AQQ}
\end {equation}
 Plancherele identity
for $\varphi$ gives
 \begin {equation}
\int w_s(\vec{p})d^3p  = \int \varphi(\vec{p})
\varphi^*(\vec{p})d^3p= \int f^2(w)d^3r =1  \label {AY255}
 \end {equation}
Application of Dubois-Reimond lemma, described in appendix,
results in $f=\pm w^{\frac{1}{2}}$. From derivative Fourier
transform  \cite {Rees}
\begin {equation}
<\vec{p}>=\int \vec{p} \varphi(\vec{p}) \varphi^*(\vec{p})d^3p=
-i\int\eta^* \frac{\partial \eta}{\partial \vec{r}}d^3r
 =\int w\,{\rm{grad}}\,
 \,{\Lambda_{\it\vec{r}}} \,  d^3r
                   \label {AY55}
\end {equation}
Then from $<\vec{p}>=\mu<\vec{u}>$ and eq.(\ref{AX25}) it follows
$$
\vec{v}=\frac{1}{\mu}\rm{grad}\,\Lambda_{\it\vec{r}}
$$
so that $\Lambda_{\vec {r}}$ is independent of $w$. Hence
analogously to $1$-dimensional case one can suppose that
$\vec{v}$ isn't fundamental DF, but  is derivative of other
fundamental DF $\gamma(\vec{r})=\Lambda_{\vec {r}}$. From that
  standard QM ansatz for  $m$ state on R$^3$ follows:
$\eta=w^{\frac{1}{2}}\exp(i\gamma)$.
 Thus, $\eta(\vec{r})$ is vector (ray)
in complex Hilbert space $\cal H$ and describes the  set
of $m$ pure states.


%

 Assuming that $\eta$  evolution is linear,
 for free $m$
evolution its operator $\hat{H}_0$ should be  even polinom of
the form
\begin {equation}
   \hat {H}_0=
    - \sum\limits_{l=1}^n b_{2l} \frac{\partial^{2l} }{\partial \vec{r}^{2l}}
                                                              \label {AU9}
 \end {equation}
 If potential field influence can be described by the addition of
real, differentiable function $U(\vec{r},t)$ to $\hat{H_0}$,
 so that
$$
    i\frac{\partial \eta}{\partial t}=\hat{H}\eta= (\hat{H}_0+U) \eta.
$$
 Then, analogously to 1-dimensional ansatz,
 the term $\partial_t w^{\frac{1}{2}}$ can be extracted from this equation and
expressed via corresponding $w,\gamma$
 $\,\vec r$-derivatives. From their term by term
comparison with corresponding $\hat H\eta$ $\,\vec r$-derivatives,
%
the  Schroedinger equation follows for $\eta$ evolution,
 at least for $\eta \in \rm{C}^4(\rm{R}^3\times[\it{t_0,T}])$..


In our initial approach, $m$ state evolution should be described
by evolution equations for $w$ and $\vec v$, but  they become
equivalent to  evolution equation for $\eta$, which unambiguously
defines $w, \vec v$. Hence it's possible to replace initial CP
for $w, \vec v$ for $m$ states and exploit CP for $\eta$
functions: $\eta\in \rm{C}^2(\rm{R}^3\times[\it{t_0,T}])$.
To make  $m$ state $\eta(\vec{r},t)$ unambiguous, the additional
constraint should be imposed on $\gamma(\vec{r})$, namely, for
any closed loop $l$ in $\rm R^3$
 \begin {equation}
   \oint_l \frac{\partial\gamma}{\partial \vec{r}}  d\vec{l}=\frac{2\pi}{\mu}n_l \label {U666}
 \end {equation}
Normally, $n_l=0$, but if the node or nodal region is located
inside loop $l$, then $n_l=0, \pm1, \pm2,...$, etc., can appear
\cite {Gh}. We consider here only the stationary noding regions
in form of infinite lines,  on such line $\vec v$ will be
undefined. The example is $2p_1, 2p_{-1}$ states of hydrogen atom
where nodal region is the  line of $z-$axis, $n_l=\pm 1$.
Description of $m$ evolution for other nodal regions types will be
considered elsewhere. It's possible to extend such formalism on
$m$ evolution in arbitrary open set $\Omega\in
\rm{C}^2(\rm{R}^3\times[\it{t_0,T}])$, it particular, it can
consist of several disjoint components $\Omega = \cup \Delta_i$,
where $\{\Delta_i \}$ are open subsets. In this case,
$\gamma(\vec{r})$ can have spatial breaks between such components.
 Thereon, it can be supposed that our formalism can exploit CP of standard
QM formalism \cite {Jauch,Vlad}.

In this formalism, $\gamma(\vec{r})$ defined up to indefinite
constant $c_\gamma$, such uncertainty contradicts to standard
geometric framework.
 To avoid it,
  $\gamma(\vec{r})$
can be replaced in $m$ state description by  dynamical
correlation $\kappa(\vec{r}_1,\vec{r}_2)$ defined as
$$
 \kappa(\vec{r}_1,\vec{r}_2)=\gamma(\vec{r}_1)-\gamma(\vec{r}_2).
$$
so that $\kappa$ doesn't depend on $c_{\gamma}$, and is bilocal
geometric object.
Hence for  arbitrary  $\vec{r}_1 \neq \vec{r}_2  $
 $$
 v(\vec{r}_1)=\frac{1}{\mu}\frac{\partial \kappa(\vec{r}_1,\vec{r}_2)}{\partial \vec{r}_1}.
$$
so that $\kappa$ can be treated as $\vec v$ generating function.
It's notable that in standard QM,  $\kappa(\vec{r}_1,\vec{r}_2)$
correlation appears in the description of quantum states by
density matrixes,  i.e. the positive, trace one operators on
$\cal H$ \cite {Jauch}. Namely, the density matrix of pure state
is equal to
\begin {equation}
   \rho(\vec{r}_1,\vec{r}_2)=
   [w(\vec{r}_1)w(\vec{r}_2)]^{\frac{1}{2}}e^{i\kappa(\vec{r}_1,\vec{r}_2)}  \label {BYB22}
\end {equation}
so that $\rho(\vec{r},\vec{r})=w(\vec{r})$. For arbitrary
observable $Q$ it gives $\bar{Q}=\rm{Tr}$$\hat{Q}\rho$; $\rho$
evolution obeys to Lioville equation
 $\dot{\rho}=[\hat{H},\rho]$
which is formally bilocal.
  In standard QM, the density matrix is considered, in
fact, as auxilary object exploited mainly for the description of
mixed states and statistical ensembles. Alternatively, the state
space of algebraic QM formalism is   density matrix space,
which considered as fundamental states  \cite {Jauch}. Our
analysis indicates that in geometric QM formulation the density
matrixes of pure states can be treated as fundamental dynamical
representation of $m$ state $|\zeta\}$. Therefore,
observational $|\zeta \}$ representation becomes
$\varrho^o_{\kappa}=\{w(\vec{r}_1),
\kappa(\vec{r}_1,\vec{r}_2)\}$. Yet,  for pure states
$\rho(\vec{r}_1,\vec{r}_2)=\eta(\vec{r}_1)*\bar{\eta}(\vec{r}_2)$, hence
$\eta(\vec{r})$ can be used  in place of $\rho(\vec{r}_1,\vec{r}_2)$
 whenever it doesn't violate described conditions. The
similar properties possess the pure states on projective Hilbert
space, in particular, such state ansatz doesn't depend on
$\gamma(\vec{r})$  component $c_{\gamma}$,  and the state
evolution is also described by bilocal equation \cite {Kib}.

  The conditions of QM
 linearity initially were formulated by Wigner \cite {Wig}, however, they are
extensively  discussed up to now \cite {Tor}. Recently,
    Jordan have shown that they are essentially
 weaker than Wigner theorem asserts  \cite {Jor}.
 Namely, if the following two conditions are fulfilled:

  i) evolution operator
 maps the set of  all pure states one to one onto itself

 ii)  for  arbitrary
  mixture of orthogonal states $\rho(t)=\sum P_i(t)\rho_i(t)$
   all $P_i$ are constant.

  It follows then that such evolution is
  linear. Both these conditions are in  good correspondence with
  geometric framework generic for FM formalism.
 Really, it was shown that the pure states $|\zeta\}$  describe
  the evolution of geometric object FP $m$ from its initial to
  final state. Plainly, such evolution shouldn't result in probabilistic
  mixture of pure states. It's also natural to assume that such evolution
   is unambiguous and reversible.
   Another arguments against QM nonlinearity involve the causality
   violations, in particular, superluminal signaling
   can occur in such theory for multiparticle systems \cite {Gisin}.

  \section {Discussion}

 Planck constant $\hbar=1$ in this formalism,
  but the same value ascribed to it in
Lorentz-Heaviside (relativistic) unit system, in which velocity of
light $c=1$. In FM framework, $\hbar$  connects particle
$\vec{r}$, $\vec p$  scales and doesn't have any additional
meaning. The superposition principle doesn't need to be
postulated separately in such formalism. Rather, as follows from
our results, the sum of two physical $m$ states $\eta_{1,2}$ with
proper complex coefficients $a_{1,2}$ can be considered as the
physical $m$ state also. In our approach, the state space is
defined by the underlying geometry and corresponding dynamics
i.e. is derivable concept. For  states
 of nonrelativistic particle $m$
 it was found to be equivalent to Hilbert space $\cal H$. However,
  for other
systems the resulting state space supposedly can  differ from
$\cal H$ analogously to algebraic QM where the state space is
defined by the observable algebra \cite {Jauch}.
The flow velocity $\vec{v}(\vec{r})$ isn't particle observable,
but its value be  consistently defined as the ratio of
$\vec{J}(\vec{r}),w(\vec{r})$ expectation values \cite {Aha2};
here $w(\vec{r})$ observable is described by the projection
operator $\hat {\Pi}(\vec{r})$; observable $\vec{J}(\vec{r})$
defined in \cite {Aha2,Schiff}. The particle evolution in QM in
some aspects is similar
 to the  motion of continuous media,
 this analogy is exploited in
hydrodynamical QM model  \cite {Gh}.
 In its axiomatic,
   Schroedinger equation replaced by Madelung equations,  it
  was claimed that they give equivalent description fo quantum systems. However, it was shown lately
 that hydrodynamic model results differ from standard QM for
  state evolution on disjoint sets,
   and so it can't give universal description
of quantum systems \cite {Wall}. In FM formalism,  FP density
evolution also described by flow continuity equation, but despite
some formal similarity, the basic theory premises are principally
different.
 In standard QM the evolution equations are or postulated $ad\, hoc$ or
derived assuming Galilean invariance of system states \cite
{Jauch,Schiff}. In FM the Schroedinger equation for massive
particle was derived assuming only space-time shift and rotational
invariance which are essentially weaker assumptions. Meanwhile,
it's well known that Galilean invariance can be derived from
Schroedinger equation, if a reference frame is associated with
free system $S$ of mass $M_s \to \infty$,  which initial state
$\Psi(t_0)$ is the wave packet of half-width $\sigma_s\to 0$ \cite
{Schiff,Aha}.





In this paper, it was shown that CFG formalism can be considered as
the mathematical basis for the consistent description of quantum
particle dynamics.
 Novel features of considered formalism can be revealed in most simple way
 from the comparison with Schroedinger QM formalism. From the formal side,
standard QM exploits two fundamental structures of different
nature: the space-time manifold $\rm{R^3}*T$ and function space
$\cal H$ defined on $\rm R^3$. In distinction, minimal FM
formalism involves only one basic structure -  fuzzy manifold
$\rm\tilde{R}^3*T$, nonrelativistic  particles  are
$\rm\tilde{R}^3$ elements - fuzzy points. Their evolution induces
the physical states which are equivalent to $\cal H$ Dirac
vectors. Resulting dynamics of massive particles described by
Shroedinger equation, which is the basis of universal quantum
dynamics.
 Quantum-classical transition in such theory is essentially more
simple than in standard QM, it's just the transition from
$\rm\tilde{R}^3$ fuzzy manifold to R$^3$ manifold, on which the
classical particles correspond to material  points  $\vec{r}(t)$.
Here only  evolution of nonrelativistic particle was considered
in FM formalism, but all QM formalism is, in fact, based on its
analysis, and there are no obvious obstacles, which forbid to
construct universal quantum dynamics in such approach.



It seems that FM approach is based on essentially more natural
axiomatic than standard QM. Its main postulate is the principal
uncertainty of some system parameters, like coordinate or
momentum, such uncertainty has geometric origin and isn't
connected $a\, priory$ with wave-like system properties. Rather,
in FM the wave-like system evolution is stipulated by these
geometric features  which described by CFG formalism. It's notable
that at fundamental level its main features related to manifold
ordering structure \cite {Zee,Schr}. It indicates that topology
of physical space-time can differ from topology of Euclidian or
Minkowski spaces and to be comparatively more weak.

Plainly, general relativity is essentially geometric theory,
meanwhile, up to now  attempts to quantize gravity meet
 serious difficulties even at axiomatic level. Hence, if consistent
geometric QM formalism will be constructed , it can help, in
principle, to develop quantum theory of gravity.
 Currently, the main impact of QM geometrization
studies is done
  on the exploit of Hilbert manifolds (\cite {Mar} and refs. therein), however,
   the results obtained in this approach have quite abstract form, and their
  applicability  to particular physical problems isn't obvious.
  Considered FM formalism possesses
simple and logical axiomatics which origin is basically
geometrical, hence
it can become the appropriate part of QM geometrization program,
 its implications can be important also for study of QM
foundations.

 \section *{Appendix}


Here the proof of Du Bois-Reimond lemma is considered according to
\cite {Lav}, it extended also for functions defined on R$^3$.

Lemma: if for continuous function $N(x)$  and arbitrary
continuous, differentiable function $\tau(x)$ for which
 $\tau(a)=\tau(b)=0$, ($a=-\infty, b=\infty$ permitted)
 and $\tau'= {\partial_x \tau}$
 $$
        I(\tau)=\int_a^b N(x)\tau' dx=0
 $$
  then $N(x)$ is constant on $[a,b]$ interval.
 Suppose that on the opposite, $N(x)$ varies on $[a,b]$, then it should be at least
two points $c_1, c_2$ for which $N$ values  differ, for example,
$N(c_2) < N(c_1)$. Let's $d_1, d_2$ to be the numbers for which
the following inequality holds
$$
N(c_2) < d_2 < d_1 < N(c_1)
$$
For large enough $n$ it's always possible to construct the
nonintersecting  intervals $[x_0, x_0+\pi/n], [x_1, x_1+\pi/n]$
confined in $[a,b]$ and such that inside $[x_0, x_0+\pi/n]$ the
inequality $N(x)> d_1$ holds, and inside another one: $N(x) <
d_2$. Let's chose  $\tau'(x)$ as follows:

 $\tau'=
\sin^3[n(x-x_0)]$ on $[x_0,x_0+\pi/n]$;

$\tau'= -\sin^3[n(x-x_1)]$ on $[x_1,x_1+\pi/n]$;

$\tau'=0$ on the rest of $[a,b]$.

 In that case,  function
$\tau(x)$ and its derivative are continuous.
In addition,  lemma premises suppose that $I(\tau)=0$,
but for  admitted $N$ variations in considered interval it follows
that for the same integral
 \begin {equation}
   I(\tau)= \int_a^b N(x)\tau'(x)dx > (d_1-d_2)\int_0^{\frac{\pi}{n}}    \sin^3  nx \, dx > 0
           \label {AU11}
 \end {equation}
Thus $N(x)$ should be constant on $[a, b]$ interval.

 For R$^3$  the proof  is given here only for
 functions defined on unbounded volume. For our problem, it's possible to
 substitute $\delta w(\vec{r})= \rm{div}\it{\vec{E}(\vec{r})}$,
  such $\vec E$ always exists for arbitrary $\delta w$. Namely, under these conditions
  the relation between $\delta w$ and $\vec{E}$
 described by  solution of Laplase equation, well known in electrodynamics
 \cite {Lan2}.
  Then one should prove that if for continuous scalar $N(\vec{r})$
   and any continuous vector $\vec{E}(\vec{r})$ for which $\rm{div}\it{\vec{E}}\to {\rm 0}$ for
$ |\vec{r}| \to \infty$
$$
      I_w(\vec{E}) = \int N {\rm{div}}{\vec{E}}\, { d^{3} r=0}
$$
   then $N(\vec{r})$ is constant on R$^3$.

Suppose that on the opposite, $N(\vec{r})$ varies on R$^3$, then
it should be at least two points $\vec{r}_1, \vec{r}_2$ for which
$N$ values differ, for example, $N(\vec{r}_2) < N(\vec{r}_1)$.
Let's $d_1, d_2$ to be the numbers for which the following
inequality holds
$$
N(\vec{r}_2) < d_2 < d_1 < N(\vec{r}_1)
$$
For large enough $n$ it's always possible to construct the
nonintersecting  spheres
 $S_{1,2}:\,|\vec{r}- \vec{r}_{1,2}| \le {\pi}/{2n}$,
such that inside $S_1$ the inequality $N(\vec{r})> d_1$ holds,
and inside $S_2$, $N(\vec{r}) < d_2$. Let's choose div$\vec{E}$
as follows:

 div $\vec{E}=
\cos^3(n|\vec{r}- \vec{r_{1}}|)$ in $S_1$;

div $\vec{E}= -\cos^3(n|\vec{r}- \vec{r_{2}}|)$ in $S_2$;

div $\vec{E}=0$ outside of $S_{1,2}$.

From lemma assumption  $I_w(\vec{E})=0$,
but for admitted $N$ difference in considered $S_{1,2}$ regions
it follows
$$
   I_w(\vec{E})= \int N {\rm{div}} \vec{E}\, {d^{3} r >
   (d_1-d_2)\int_0^{\frac{\pi}{2n}}}
    \cos^3 (n|\vec{r}|) \, {d^3  r > 0}
$$
Thus, $N(\vec{r})$ should be constant.



\end {document}